\documentclass{article}
\usepackage{spconf,amsmath,graphicx,bm,amssymb}

\newcommand{\be}{\mathbf{e}}

\newcommand{\bQ}{\mathbf{Q}}
\newcommand{\bc}{\mathbf{c}}

\newcommand{\bI}{\mathbf{I}}

\newcommand{\bx}{\mathbf{x}}
\newcommand{\bX}{\mathbf{X}}
\newcommand{\bY}{\mathbf{Y}}
\newcommand{\bT}{\mathbf{T}}

\newcommand{\br}{\mathbf{r}}
\newcommand{\bs}{\mathbf{s}}
\newcommand{\bS}{\mathbf{S}}
\newcommand{\bt}{\mathbf{t}}
\newcommand{\by}{\mathbf{y}}

\newcommand{\bA}{\mathbf{A}}
\newcommand{\bB}{\mathbf{B}}

\newcommand{\bH}{\mathbf{H}}

\newcommand{\bP}{\mathbf{P}}
\newcommand{\bU}{\mathbf{U}}
\newcommand{\bZ}{\mathbf{Z}}

\renewcommand{\frac}{\dfrac}

\newcommand{\T}{{{T}}}

\newcommand{\wuhao}{\fontsize{9pt}{\baselineskip}\selectfont}
\newcommand{\xiaowuhao}{\fontsize{8.5pt}{\baselineskip}\selectfont}
\newcommand{\liuhao}{\fontsize{8.15pt}{\baselineskip}\selectfont}

\newtheorem{dingli}{Theorem~}

\title{On the Equivalence of Semidifinite Relaxations for MIMO Detection with General Constellations}
\name{ Ya-Feng Liu$^{\star}$, Zi Xu$^{\dag}$, and Cheng Lu$^{\S}$}

\address{$^{\star}$LSEC, ICMSEC, AMSS, Chinese Academy of Sciences, Beijing, China \\[2pt]
    $^{\dag}$ Department of Mathematics, College of Sciences, Shanghai University, Shanghai,
    China\\[2pt]
    $^{\S}$School of Economics and Management, North China Electric Power University, Beijing, China\\[2pt]
  Email: yafliu@lsec.cc.ac.cn, xuzi@i.shu.edu.cn, lucheng1983@163.com}
\begin{document}
%
\maketitle
\wuhao {
\begin{abstract}
The multiple-input multiple-output (MIMO) detection problem is a fundamental problem in modern digital communications.
Semidefinite relaxation (SDR) based algorithms are a popular class of approaches to solving the problem because the algorithms have
a polynomial-time worst-case complexity and generally can achieve a good detection error rate performance.
In spite of the existence of various different SDRs for the MIMO detection problem in the literature, very little is known about the relationship between these SDRs. This paper aims to fill this theoretical gap. In particular, this paper shows that two existing SDRs for the MIMO detection problem, which take quite different forms and are proposed by using different techniques, are equivalent. As a byproduct of the equivalence result, the tightness of one of the above two SDRs under a sufficient condition can be obtained.

%
%
\end{abstract}
\section{Introduction}
\label{sec:intro}

The MIMO detection problem is a fundamental problem in modern digital communications, which has been extensively studied for several decades \cite{Yang}.
Recently, it has received renewed interest, due to its potential applications in massive MIMO technology in 5G \cite{Yang,Huikang}. The MIMO detection problem is generally modeled as a complex quadratic optimization problem. Various algorithms have been proposed to solve the problem.
One of the most celebrated algorithms is the sphere decoder algorithm \cite{Damen,Murugan}. 
The sphere decoder algorithm is a special branch-and-bound based enumeration algorithm, which is guaranteed to find the globally optimal solution of the problem. However, the worst-case and expected complexity of the sphere decoder algorithm are exponential \cite{Verdu1989,complexitysphere}. Motivated by some real-time applications, some efficient sub-optimal algorithms have also been proposed. For instance, the zero-forcing detector algorithm \cite{Schneider}, the minimum mean-squared error detector algorithm \cite{Honig}, and the decision feedback detector algorithm \cite{Varanasi}, are all low-complexity sub-optimal algorithms. 
The performance of these algorithms are generally not good in the sense that their detection error rates are very high.

In the past two decades, the semidefinite relaxation (SDR) detector algorithms have received great attention \cite{Tan}--\cite{So2010}. The SDR detector has been proposed first for the BPSK constellation \cite{Tan,Ma} and then extended to the QPSK constellation \cite{Ma2004,Steingrimsson}.
It has been shown that the SDR detector achieves a considerably lower detection error rate than all previously mentioned sub-optimal algorithms.
Moreover, the SDR detector has a guaranteed polynomial-time worst-case complexity.
To understand why the SDR detector performs remarkably well in practice, the approximation ratios of some SDR based algorithms have been
studied in \cite{So2007,Zhang,Kisialiou}. In particular, for the BPSK case, it has been shown in \cite{Jalden2008} that the SDR based algorithm can achieve the maximum possible diversity order. In addition to the above analysis results, some sufficient conditions, under which the SDRs are tight, have been also identified in \cite{Jalden,So2010}.

Besides the BPSK and QPSK cases, the SDR based algorithms have also been extended to other general constellation cases, especially the high-order QAM and $M$-PSK constellations \cite{Ma2004b,Sidiropoulos2006}. Various SDR models have been proposed. For example,
in \cite{Mobasher}, the detection problem is first reformulated as a quadratic integer optimization problem, and then some SDRs are designed by exploiting the special structure of the quadratic integer optimization problem. In \cite{Fan}, the number of design variables in the above quadratic integer optimization reformulation is further reduced, and a more compact SDR is proposed. Very recently, an SDR for the general $M$-PSK constellation is proposed in \cite{Lu}, which is an enhanced SDR over the classical complex SDR and is obtained by
adding valid linear cuts into an equivalent real reformulation of the classical complex SDR. Numerical results in \cite{Lu} show that the enhanced SDR is much tighter than the classical complex SDR.

While various SDRs have been proposed for the MIMO detection problem due to different motivations and/or by using
different techniques, there are very few works studying the relationship between these SDRs. To the best of our knowledge,
the only work along this line is \cite{Ma2009}, where some SDRs for the QAM constellation have been compared.
The goal of this work is to provide a comprehensive comparison of existing SDRs for the MIMO detection problem with a general constellation.
Due to the space limitation, we only present one of our main results here (and more results will be presented in the journal extension).
In particular, we show that an enhanced SDR proposed in \cite{Lu} and a famous SDR proposed in \cite{Mobasher} are equivalent; see (E$\mathbb{R}$SDR1) and (E$\mathbb{R}$SDR2) further ahead.
As a byproduct of the above equivalence result, we can show the tightness of (E$\mathbb{R}$SDR2) proposed in \cite{Mobasher} under a sufficient condition. This tightness result remains unknown until this paper.

We adopt the following standard notations in this paper. We use $\mathbb{C}^{m\times n}$ ($\mathbb{R}^{m\times n}$) and $\mathbb{C}^{m}$ ($\mathbb{R}^{m}$) to denote the set of $(m\times n)$-dimensional complex (real) matrices and $m$-dimensional complex (real) vectors, respectively. We use $(\cdot)^{\T}$ and $(\cdot)^{\dagger}$ to denote the transpose and Hermitian transpose of a matrix/vector, respectively. We use $\mathrm{Re}(\cdot)$ and $\mathrm{Im}(\cdot)$ to denote the element-wise real and imaginary parts of a complex matrix/vector/number, respectively. We use $\|\cdot\|_2$ and $\|\cdot\|_{\infty}$ to denote the $2$-norm and $\infty$-norm of a matrix/vector. The notations $\be,~\bm{0},$ and $\bI$
represent the all-one vector, the all-zero matrix/vector, and the identity matrix of appropriate sizes, respectively. For a given complex number $x,$  $\arg{(x)}$ denotes its argument. For a given vector $\bt,$ $\text{Diag}(\bt)$ denotes the diagonal matrix formed by it. Finally, for two given matrices $\bA$ and $\bB$ (of appropriate sizes), $\bA\succeq \bm{0}$ means that $\bA$ is a Hermitian positive semidefinite (PSD) matrix; $\bA\bullet \bB$ denotes the trace of their product $\bA\bB$, i.e., $\sum_{i}\sum_{j}A_{i,j}B_{j,i}$; and $\bA\otimes\bB$ denotes their Kronecker product.

\section{MIMO Detection Problem Formulation}

Consider a complex-valued MIMO channel model \begin{equation}\label{rHv}\br=\bH\bx^{\ast}+\bm{\nu},\end{equation} where $\br \in\mathbb{C}^{m}$ is
the vector of received signals, $\bH\in\mathbb{C}^{m\times n}$ is an
$m\times n$ complex channel matrix (for $n$ inputs and $m$ outputs with $m\geq n$), $\bx^{\ast}\in\mathbb{C}^{n}$ is the vector of transmitted
symbols, and $\bm{\nu}\in\mathbb{C}^{m}$ is an additive {white circularly symmetric Gaussian noise} with zero mean. 
Throughout the paper, we assume that the $M$-PSK modulation scheme with $M\geq 2$ is adopted\footnote{The main results in this paper can also be extended to the QAM case.}. Then, each entry $x_i^{\ast}$ of $\bx^{\ast}$ belongs to a finite set of symbols
$$\left\{\exp(\textbf{i}\theta)\mid\theta=\frac{2(j-1)\pi}{M},~j=1,2,\ldots,M\right\},~i=1,2,\ldots,n,$$ where $\textbf{i}$ is the imaginary unit (which satisfies $\textbf{i}^2=-1$).
The MIMO detection problem is to recover the vector of transmitted symbols $\bx^{\ast}$ from the vector of received signals $\br$ based on the knowledge of the channel matrix $\bH$.
The mathematical formulation of the problem is 
\begin{equation*}\begin{array}{cl}
\displaystyle \min_{\bx \in \mathbb{C}^{n}} & \left\|\bH\bx-\br\right\|_2^2 \\[3pt] \tag{P}
\mbox{s.t.} & |x_i|^2=1,~\arg{(x_i)}\in \mathcal{A},~i=1,2,\ldots,n,
\end{array}\end{equation*}
where $\mathcal{A}=\left\{0, 2\pi/M,\ldots,2(M-1)\pi/M\right\}.$

\section{Review of Some Existing SDRs for (P)}

The MIMO detection problem (P) is NP-hard \cite{Verdu1989}. Therefore, there is no polynomial-time algorithms which can solve it to global optimality in general (unless P=NP). In the last two decades, the SDR based algorithms have
been widely studied in the signal processing and wireless communication community \cite{Luo2010,Pu2018} and particularly have been designed for solving problem (P).
The SDR based algorithms for solving problem (P) not only enjoy a polynomial-time worst-case complexity but also generally achieve a very good detection error rate performance. In this section, we
briefly review some existing SDRs for problem (P).

For notational simplicity, 
let $\bQ=\bH^{\dag} \bH$ and $\bc=-\bH^{\dag} \br;$ let $\bs=[s_1, s_2, \ldots, s_M]^T\in \mathbb{C}^{M}$ be the vector of all constellation symbols, where $$s_j = \cos\left(\frac{2(j-1)\pi}{M}\right) + \textbf{i} \sin\left(\frac{2(j-1)\pi}{M}\right),~j=1,2,\ldots,M;$$ and finally let $\bs_R=\mathrm{Re}(\bs)$ and $\bs_I=\mathrm{Im}(\bs).$

By introducing an $n\times n$ complex matrix $\bX=\bx\bx^\dagger$, problem (P) can be equivalently reformulated as
\begin{align*}
\min_{\bx,\bX} ~&~ \bQ\bullet \bX+2\mathrm{Re}\left(\bc^{\dagger}\bx\right) \\
\mbox{s.t.~} ~&~ X_{i,i} = 1,~i=1,\ldots,n,\\
&~\arg\left(x_i\right)\in \mathcal{A},~i=1,\ldots,n,\\
&~ \bX=\bx\bx^\dagger,
\end{align*}where the variables $\bx \in \mathbb{C}^{n}$ and $\bX \in \mathbb{C}^{n\times n}$ and $X_{i,i}$ is the $i$-th diagonal entry of $\bX.$ A straightforward (but loose) SDR of problem (P) is \begin{align*}
\min_{\bx,\bX} ~&~ \bQ\bullet \bX+2\mathrm{Re}\left(\bc^{\dagger}\bx\right) \\ \tag{$\mathbb{C}$SDR}
\mbox{s.t.~} ~&~  X_{i,i} = 1,~i=1,\ldots,n,\\
&~ \bX\succeq \bx\bx^\dagger,
\end{align*} which drops the argument constraints
$\arg\left(x_i\right)\in \mathcal{A}$ for all $i=1,2,\ldots,n$ and relaxes the nonconvex constraint
$\bX=\bx\bx^\dagger$ to $$\bX\succeq \bx\bx^\dagger\Longleftrightarrow \begin{bmatrix}
1 &\bx^{\T}\\
\bx &\bX\\
\end{bmatrix}
\succeq \bm{0}.$$  It has been shown in \cite{Lu} that ($\mathbb{C}$SDR) is equivalent to the following real SDR
\begin{equation}\begin{array}{cl}
\displaystyle \min_{\by,\bY} & \hat{\bQ} \bullet \bY +2\hat{\bc}^{\T}\by\\[3pt]
\mbox{s.t.} & Y_{i,i}+Y_{n+i,n+i}=1,~i=1,2,\ldots,n, \\[5pt]
& \bY\succeq\by\by^{\T},\tag{$\mathbb{R}$SDR}
\end{array}\end{equation}
where the variables $\by\in \mathbb{R}^{2n}$ and $\bY\in \mathbb{R}^{2n\times 2n}$ and
\begin{align}\label{Qcy}
\hat{\bQ}=\begin{bmatrix}
\textrm{Re}(\bQ) ~&-\textrm{Im}(\bQ)\\[3pt]
\textrm{Im}(\bQ) ~&\textrm{Re}(\bQ)\\
\end{bmatrix},~ \hat{\bc}=\begin{bmatrix} 
\textrm{Re}(\bc)\\[3pt]
\textrm{Im}(\bc) \\
\end{bmatrix},~\by=\begin{bmatrix} 
\textrm{Re}(\bx) \\[3pt]
\textrm{Im}(\bx)\\
\end{bmatrix}.
\end{align}

Based on ($\mathbb{R}$SDR), an enhanced SDR for (P) has recently been proposed in \cite{Lu}. Define the following $3\times 3$ matrices
\begin{equation}\label{Yi}{\bY}_i=\begin{bmatrix}
1 ~&y_{i} ~&y_{n+i}\\[3pt]
y_{i} ~&Y_{i,i} ~&Y_{i,n+i}\\[3pt]
y_{n+i} ~&Y_{n+i,i} ~&Y_{n+i,n+i}\\
\end{bmatrix},~i=1,2,\ldots,n\end{equation}
and \begin{equation}\label{Pk}\bP_j=  \begin{bmatrix}
1 \\[3pt]
\textrm{Re}(s_j)\\[5pt]
\textrm{Im}(s_j)
\end{bmatrix}
\begin{bmatrix}
1 & \textrm{Re}(s_j) & \textrm{Im}(s_j)
\end{bmatrix},~j=1,2,\ldots,M.\end{equation} By the definition of $\by$ in \eqref{Qcy}, ideally each ${\bY}_i$ in \eqref{Yi} must be one of matrices
$\bP_j$ with $j=1,2,\ldots,M,$ i.e., $${\bY}_i\in\left\{\bP_1,~\bP_2,\ldots,\bP_{M}\right\},~i=1,2,\ldots,n.$$
By relaxing the above combinatorial constraints and dropping some redundant constraints, reference \cite{Lu} proposes the following enhanced SDR for (P):
{\xiaowuhao{\begin{equation}
\begin{array}{cl}
\displaystyle \min_{\by,\bY,\bt} &  \hat{\bQ} \bullet \bY + 2\hat{\bc}^{\T}\by\\[3pt]
\mbox{s.t.} & \displaystyle {\bY}_i=\sum_{j=1}^{M} t_{i,j} \bP_j,~i=1,2,\ldots,n,\\[3pt] \tag{E$\mathbb{R}$SDR1}
& \bA\bt = \be_n,~\bt\geq\bm{0},\\[3pt]
& \bY\succeq\by\by^{\T},
\end{array}
\end{equation}}}where the variables $\by\in \mathbb{R}^{2n},$ $\bY\in \mathbb{R}^{2n\times 2n},$ $\bt\in \mathbb{R}^{Mn},$ $\hat \bQ$ and $\hat \bc$ are defined in \eqref{Qcy}, ${\bY}_i$ is defined in \eqref{Yi}, $\bP_j$ is defined in \eqref{Pk}, and
$$\bS = \bI_n \otimes \bs^{\T},~\bA = \bI_n \otimes \be_M^{\T}.$$ In (E$\mathbb{R}$SDR1), $\bt=[\bt_1^{\T}, \bt_2^{\T}, \ldots, \bt_n^{\T}]^T$ and $\bt_i=[t_{i,1},t_{i,2},\ldots, t_{i,M}]^{\T}\in \mathbb{R}^{M}.$ Due to the symmetry of ${\bY}_i,$ the constraint ${\bY}_i=\sum_{j=1}^{M} t_{i,j} \bP_j$ can be explicitly expressed as the following $5$ linear constraints:
\begin{equation}\label{linearcons}
\begin{array}{rl}
\displaystyle y_i=\displaystyle\bt_i^{\T}\bs_R,~y_{n+i}=\bt_i^{\T}\bs_I,~Y_{i,i}=\displaystyle\bs_R^{\T}\text{Diag}(\bt_i)\bs_R, \\[12pt]
\displaystyle Y_{n+i,n+i}=\bs_I^{\T}\text{Diag}(\bt_i)\bs_I,~Y_{i,n+i}=\displaystyle\bs_R^{\T}\text{Diag}(\bt_i)\bs_I.
\end{array}
\end{equation} 

Another interesting SDR for problem (P) is proposed in \cite{Mobasher} based on the following observation: for each $x_i^{\ast}$ of $\bx^{\ast},$ there holds $x_i^{\ast} = \bt_i^{\T} \bs,$
where only one entry of $\bt_i\in \mathbb{R}^{M}$ is one and all the others are zero. Then, problem (P) is reformulated in \cite{Mobasher} as follows:
\begin{align*}
\min_{\bt} ~&~ \bt^T \bar \bQ \bt+2 \bar \bc^{\T}\bt \\
\mbox{s.t.~} ~&~ \bA\bt = \be_n,~\bt\geq\bm{0},\\
&~\bt \in \left\{0, 1\right\}^{Mn},
\end{align*} where \begin{equation}\label{hatS}\bar \bQ =   \hat \bS^{\T}\hat \bQ \hat \bS,~\bar \bc = \hat \bS^{\T} \hat \bc,~\text{and}~\hat \bS = \left[
                                \begin{array}{c}
                                  \textrm{Re}(\bS) \\
                                  \textrm{Im}(\bS) \\
                                \end{array}
                              \right].\end{equation} 
Based on the above reformulation and by exploiting the special structure of vector $\bt,$ reference \cite{Mobasher} proposes the following SDR\footnote{A slight difference between (E$\mathbb{R}$SDR2) presented here and (Model III) in \cite{Mobasher} lies in the elimination of one variable in each $\bt_i$ by using the property that the summation of $\bt_i$ is equal to one for $i=1,2,\ldots,n.$} 
\begin{align*}
\min_{\bt,\bT} ~&~ \bar \bQ \bullet \bT +2 \bar \bc^{\T}\bt \\ \tag{E$\mathbb{R}$SDR2}
\mbox{s.t.~} ~&~ \bA\bt = \be_n,~\bt\geq\bm{0},\\
             ~&~ \bT\succeq \bt\bt^{\T},\\
             ~&~ \bT_{i,i} = \text{Diag}(\bt_i),~i=1,2,\ldots, n,
\end{align*} where the variables $\bt\in \mathbb{R}^{Mn}$ and $\bT\in \mathbb{R}^{Mn\times Mn}$ and $\bT_{i,i}\in \mathbb{R}^{M\times M}$ denotes the $i$-th diagonal block of matrix $\bT.$
The last constraint $\bT_{i,i} = \text{Diag}(\bt_i)$ requires that $\bT_{i,i}$ is a diagonal matrix and all its diagonal entries are equal to $\bt_i.$

\section{Main Results}
In this section, we present the main result of this paper. We show, somewhat surprisingly, that (E$\mathbb{R}$SDR1) and (E$\mathbb{R}$SDR2) are equivalent,
although they are derived by using different techniques and due to different motivations and they take quite different forms. The equivalence here means
that, for any given feasible point $(\bT, \bt)$ of (E$\mathbb{R}$SDR2), there exists a feasible point $(\by, \bY, \bt)$ of (E$\mathbb{R}$SDR1) such that the two problems have the same objective value at the corresponding points; and for any given feasible point $(\by, \bY, \bt)$ of (E$\mathbb{R}$SDR1), there exists a feasible point $(\bT, \bt)$ of (E$\mathbb{R}$SDR2) such that the two problems also have the same objective value.

\begin{dingli}\label{thmequivalence}
  (E$\mathbb{R}$SDR1) and (E$\mathbb{R}$SDR2) are equivalent. 
\end{dingli}
\textbf{Proof:} Due to the space reason, we only give a proof outline here. To show the theorem, it suffices to show that a pair of the feasible points of (E$\mathbb{R}$SDR1) and (E$\mathbb{R}$SDR2) satisfies the following relationship
\begin{equation}\label{connection}
  \hat \bS \bT  \hat \bS^{\T} = \bY~\text{and}~\hat \bS \bt = \by,
\end{equation} where $\hat\bS$ is given in \eqref{hatS}. The conditions in \eqref{connection} guarantee that the two SDRs have the same objective value.
Now, given any feasible point $(\bT, \bt)$ of (E$\mathbb{R}$SDR2), one can easily check that the same $\bt$ jointly with $\by$ and $\bY$ given in \eqref{connection} is a feasible point of (E$\mathbb{R}$SDR1) and they achieve the same objective value as that of (E$\mathbb{R}$SDR2) at $(\bT, \bt)$. Next, given any feasible point $(\by, \bY, \bt)$ of (E$\mathbb{R}$SDR1), we shall construct a matrix $\bT$ such that it, jointly with the given $\bt,$ is a feasible point of (E$\mathbb{R}$SDR2) and the two problems have the same objective value at these two points.

Without loss of generality, suppose that the PSD matrix $\bY-\by\by^{\T}$ is not zero. 
Let $r\geq 1$ denote the rank of $\bY-\by\by^{\T}.$ Furthermore, suppose $\bY-\by\by^{\T}$ has the following eigenvalue decomposition $\bY-\by\by^{\T} = \bU \bm{\Lambda} \bU^{\T},$
where $\bU\in \mathbb{R}^{2n\times r}$ and $\bm{\Lambda} \succ \bm{0}.$
Similarly, for each $i=1,2,\ldots,n,$ one can easily show that $\text{Diag}(\bt_i) -\bt_i\bt_i^{\T}$ is PSD due to the fact that $\be_M^{\T}\bt_i =1$ and $\bt_i\geq \bm{0}.$
Suppose that $\text{Diag}(\bt_i) -\bt_i\bt_i^{\T} = \bU_i \bm{\Lambda}_i \bU_i^{\T},$
where $\bU_i\in \mathbb{R}^{M\times M}$ and $\bm{\Lambda}_i \succeq \bm{0}.$
Construct the vectors
\begin{equation*}\label{eta}\bm{\eta}_{i}=\left\{\begin{array}{@{}ll}
\bm{\Lambda}_i^{1/2}\bU_i^T \bs_R,~~\textrm{for}~ i=1,2,\ldots,n;\\[3pt]
\bm{\Lambda}_i^{1/2}\bU_i^T \bs_I,~~\textrm{for}~ i=n+1,n+2,\ldots,2n.
\end{array}\right.\end{equation*}
By using \eqref{linearcons}, one can check that the above $\left\{\bm{\eta}_i\in\mathbb{R}^{M\times 1}\right\}$ satisfy
\begin{equation}\label{etanorm}
\begin{array}{rl}
  \left\|\bm{\eta}_{i}\right\|^2=Y_{i,i}-y_i^2,~i=1,2,\ldots,2n,\\[5pt]
  \bm{\eta}_{i}^T\bm{\eta}_{n+i}=Y_{i,n+i}-y_iy_{n+i},~i=1,2,\ldots,n.
\end{array}
\end{equation} 

Suppose
$\bY-\by\by^T= \left[
              \begin{array}{ccc}
                \bm{\xi}_1  & \ldots & \bm{\xi}_{2n} \\
              \end{array}
            \right]^T\left[
              \begin{array}{cccc}
                \bm{\xi}_1 & \ldots & \bm{\xi}_{2n} \\
              \end{array}
            \right],
$ where $\bm{\xi}_i\in\mathbb{R}^{r\times 1}$ for all $i.$ Obviously,
\begin{equation}\label{xiproduct}\bm{\xi}_i^T \bm{\xi}_j = Y_{i,j}-y_iy_j,~i,j=1,2,\ldots,2n.\end{equation}
One can show from \eqref{etanorm} and \eqref{xiproduct} that there exist $\left\{\bZ_i\in \mathbb{R}^{r\times M}\right\}$ such that
  \begin{equation}\label{Zi}
    \bZ_i^T \bZ_i \preceq \bI_{M},~\bZ_i\bm{\eta}_i = \bm{\xi}_i,~\bZ_i\bm{\eta}_{n+i} = \bm{\xi}_{n+i},~i=1,2,\ldots,n.
  \end{equation}

Now, we can construct the desired matrix $\bT\in \mathbb{R}^{nM \times nM}.$ Let the $(i,j)$-th block of $\bT$ be
\begin{equation*}\bT_{i,j}=\left\{\begin{array}{@{}ll}
\bt_i\bt_j^T + \bX_i^{\T}(\bY-\by\by^T)\bX_j,~~\textrm{if}~ i\neq j;\\[3pt]
 \text{Diag}(\bt_i) ,~~\textrm{if}~ i=j,
\end{array}\right.\end{equation*} where $\bX_i=\bU\bm{\Lambda}^{-1/2}\bZ_i \bm{\Lambda}_i^{1/2}\bU_i^{\T}\in \mathbb{R}^{2n \times M},~i=1,2,\ldots, n$
and $\left\{\bZ_i\right\}_{i=1}^n$ are given in \eqref{Zi}. One can check that the above constructed $\bT$ and the given $\bt$ jointly satisfy all constraints in (E$\mathbb{R}$SDR2) and equations in \eqref{connection} (and thus they achieve the same objective value as that of (E$\mathbb{R}$SDR1) at $(\by, \bY, \bt)$).  This completes the proof. \hfill {\bf Q.E.D.}

Two remarks on Theorem \ref{thmequivalence} are in order.~First, combining Theorem \ref{thmequivalence} and \cite[Theorem 4.4]{Lu}, we can immediately obtain the following tightness result of (E$\mathbb{R}$SDR2). 
\begin{dingli}\label{thmtightness}
  Suppose that $M\geq 2.$ If the inputs $\bH$ and $\bm{\nu}$ in \eqref{rHv} satisfy
  \begin{equation}\label{condition}
  \lambda_{\min}\left(\bH^{\dag}\bH\right)\sin\left(\frac{\pi}{M}\right)>\left\|\bH^{\dag}\bm{\nu}\right\|_{\infty},
\end{equation} where $\lambda_{\min}\left(\bH^{\dag}\bH\right)$ denotes the smallest eigenvalue of $\bH^{\dag}\bH,$
   then (E$\mathbb{R}$SDR2) is tight for (P).
\end{dingli}
The sufficient condition in \eqref{condition} is intuitive, which roughly says that problem (P) is an ``easy'' problem (polynomial-time solvable) if {the channel matrix} is well conditioned and {the number of constellation points} and {the noise level} are below a certain threshold.
Second, Theorem \ref{thmequivalence} reveals that there is some ``redundancy'' in (E$\mathbb{R}$SDR2).
In particular, we can see from \eqref{connection} that there is a correspondence between the feasible sets of (E$\mathbb{R}$SDR2) and (E$\mathbb{R}$SDR1) and
all information contained in the high-dimensional space $\left(\bT,\bt\right)$ in (E$\mathbb{R}$SDR2) is kept in the low-dimensional space $\left(\by,\bY,\bt\right)$ in (E$\mathbb{R}$SDR1) under the mapping in \eqref{connection}. To be more specific, the matrix variable in (E$\mathbb{R}$SDR1)
is of dimension $2n\times 2n$ but the matrix variable in (E$\mathbb{R}$SDR2) is of dimension $Mn\times Mn.$ Hence, (E$\mathbb{R}$SDR1) should be more efficiently solvable
than (E$\mathbb{R}$SDR2) especially when $M$ is much larger than $2.$ The equivalence shown in Theorem \ref{thmequivalence} provides useful insight into
possibly reducing the ``redundancy'' in existing SDRs for more general combinatorial optimization problems and designing new computationally more efficient SDRs.


\section{Simulation Results}
In this section, we present some preliminary simulation results to verify the equivalence between (E$\mathbb{R}$SDR1) and (E$\mathbb{R}$SDR2).
In our simulations, all entries of the channel matrix $\bH\in\mathbb{C}^{m\times n}$ are generated independently and identically according to the standard complex Gaussian distribution, and all entries of the transmitted symbol vector $\bx^{\ast}\in\mathbb{C}^{n}$ are drawn independently and uniformly from the $8$-PSK constellation. In our simulation, we focus on the 8-PSK constellation with $(m, n) = (10, 10)$. We define the SNR as follows: 
\begin{equation*}
\mbox{SNR}=\frac{\mathbb{E}[\|\bH \bx^{\ast}\|_2^2]}{\mathbb{E}[\|\bm{\nu}\|_2^2]}= \frac{m\sigma_{\bx}^2}{\sigma_{\bm{\nu}}^2},
\end{equation*}
where $\sigma_{\bx}^2 = \mathbb{E}[\|\bx^{\ast}\|_2^2],$ $\sigma_{\bm{\nu}}^2 = \mathbb{E}[\|\bm{\nu}\|_2^2],$ and $\mathbb{E}[\cdot]$ is the expectation operator. 
For each SNR value, we randomly generate $100$ problem instances $(\bH,\bx^{\ast}, \bm{\nu})$ and the results presented below are obtained by averaging over all generated instances. We use the solver SeDuMi in CVX \cite{cvx} to solve the two SDRs, i.e., (E$\mathbb{R}$SDR1) and (E$\mathbb{R}$SDR2).

Fig. \ref{error} shows the average difference of the optimal objective values of (E$\mathbb{R}$SDR1) and (E$\mathbb{R}$SDR2) and the average difference of the first and second equations in \eqref{connection} at the optimal solutions, i.e.,  $\|\hat \bS \bT  \hat \bS^{\T} - \bY\|_2$ and $\|\hat \bS \bt - \by\|_2$, versus different SNRs. As can be observed from Fig. \ref{error}, the difference under all these three measures is very small (in the order of $1$e$-4$) over the whole range of tested SNRs, and this shows that (E$\mathbb{R}$SDR1) and (E$\mathbb{R}$SDR2) are indeed equivalent. Fig. \ref{fig_time} shows the average CPU time taken to solve  (E$\mathbb{R}$SDR1) and (E$\mathbb{R}$SDR2) versus different SNRs. We can see clearly from Fig. \ref{fig_time} that solving (E$\mathbb{R}$SDR1) is much more efficient than solving (E$\mathbb{R}$SDR2). It is expected that the time difference of solving the two SDRs will become larger as the dimension of the problem (especially the number of constellation points) increases. All the above simulation results are consistent with our analysis.


\begin{figure}[t]
	\includegraphics[width=0.42\textwidth]{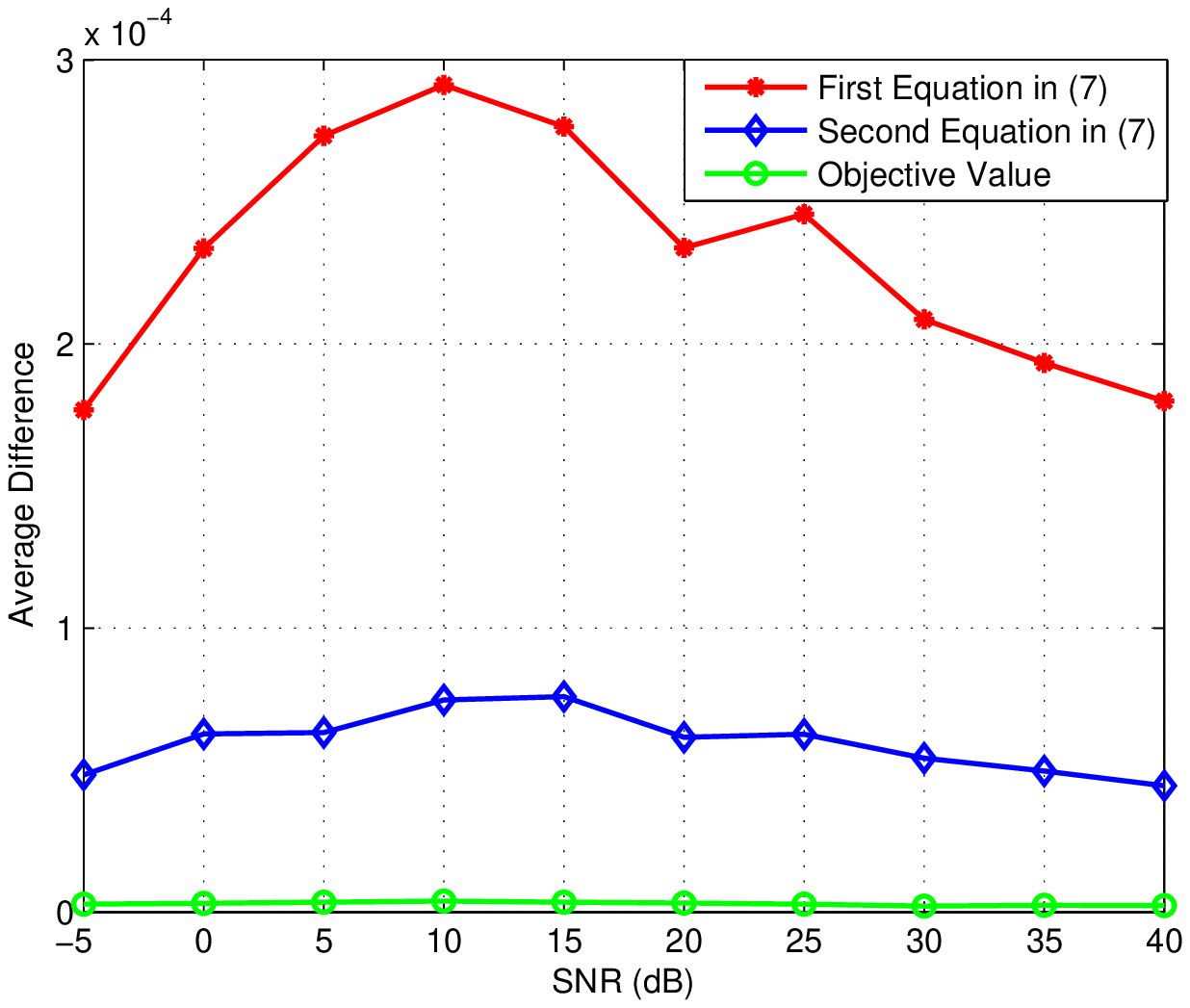}
	\caption{Average difference of solutions of (E$\mathbb{R}$SDR1) and (E$\mathbb{R}$SDR2) under different measures.}
	\label{error}
\end{figure}

\begin{figure}[t]
	\includegraphics[width=0.42\textwidth]{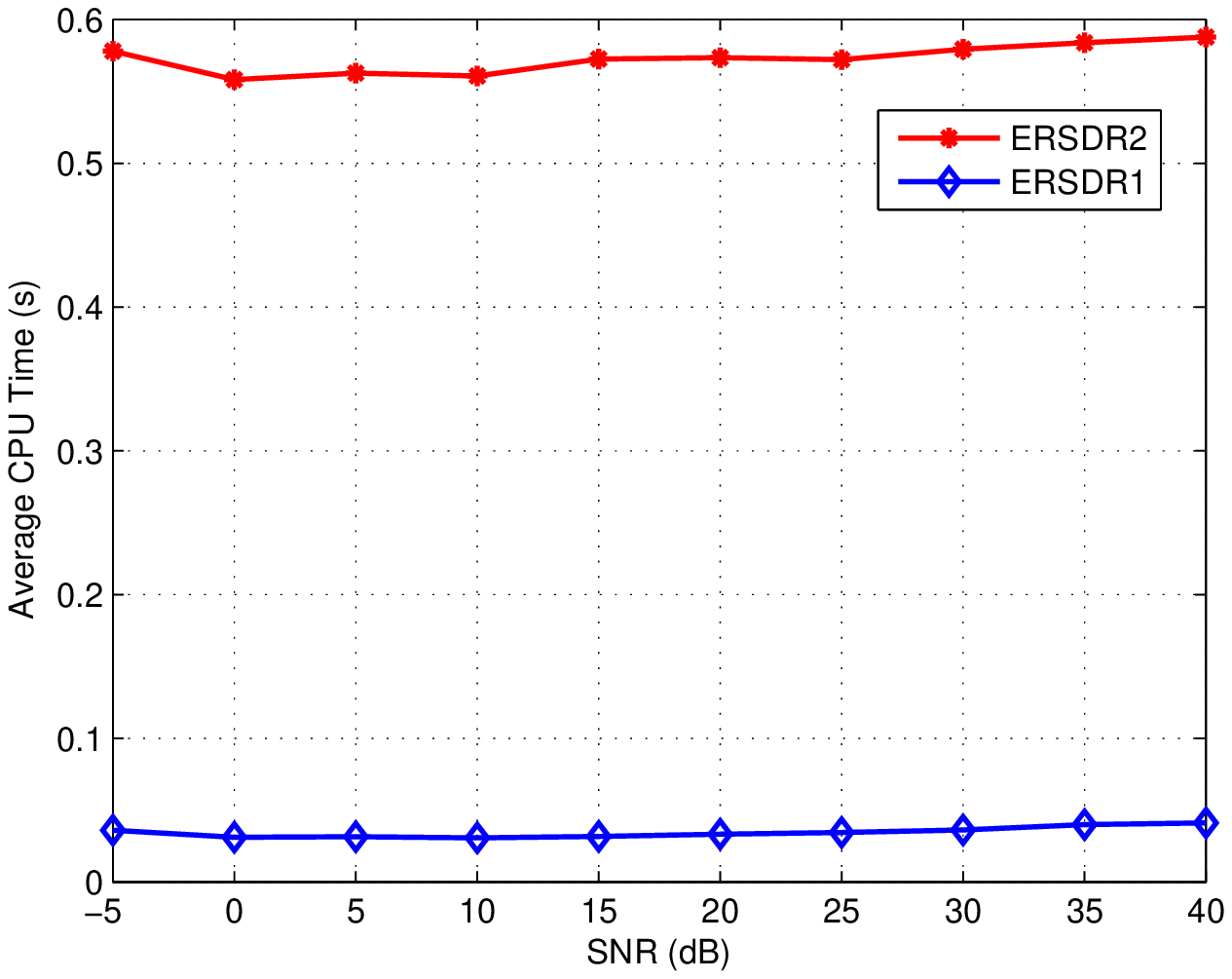}
		\caption{Average CPU time of solving (E$\mathbb{R}$SDR1) and (E$\mathbb{R}$SDR2).}
	\label{fig_time}
\end{figure}

\newpage

{}}
\end{document}